\documentclass[aps,prb,twocolumn,showpacs,superscriptaddress]{revtex4}
\usepackage[T1]{fontenc}
\usepackage[utf8]{inputenc}
\usepackage{newunicodechar}
\newunicodechar{ü}{\"{u}}
\newunicodechar{ö}{\"{o}}
\newunicodechar{Ö}{\"{O}}
\newunicodechar{ø}{\o}
\newunicodechar{á}{\'{a}}
\newunicodechar{Γ}{\Gamma}

\usepackage{graphicx}
\usepackage{hyperref}
\usepackage{graphicx}
\usepackage{color}
\usepackage[normalem]{ulem}

\begin{document}


\title{Growth and electronic structure of epitaxial single-layer WS$_2$ on Au(111)}


\author{Maciej Dendzik}
\affiliation{Department of Physics and Astronomy, Interdisciplinary Nanoscience Center, Aarhus University, 8000 Aarhus C, Denmark}
\author{Matteo Michiardi}
\affiliation{Department of Physics and Astronomy, Interdisciplinary Nanoscience Center, Aarhus University, 8000 Aarhus C, Denmark}
\author{Charlotte Sanders}
\affiliation{Department of Physics and Astronomy, Interdisciplinary Nanoscience Center, Aarhus University, 8000 Aarhus C, Denmark}
\author{Marco Bianchi}
\affiliation{Department of Physics and Astronomy, Interdisciplinary Nanoscience Center, Aarhus University, 8000 Aarhus C, Denmark}
\author{Jill A. Miwa}
\affiliation{Department of Physics and Astronomy, Interdisciplinary Nanoscience Center, Aarhus University, 8000 Aarhus C, Denmark}
\author{Signe S. Grønborg}
\affiliation{Department of Physics and Astronomy, Interdisciplinary Nanoscience Center, Aarhus University, 8000 Aarhus C, Denmark}
\author{Jeppe Vang Lauritsen}
\affiliation{Department of Physics and Astronomy, Interdisciplinary Nanoscience Center, Aarhus University, 8000 Aarhus C, Denmark}
\author{Philip Hofmann}
\affiliation{Department of Physics and Astronomy, Interdisciplinary Nanoscience Center, Aarhus University, 8000 Aarhus C, Denmark}
\email[]{philip@phys.au.dk}


\date{\today}

\begin{abstract}
Large-area single-layer WS$_2$ is grown epitaxially on Au(111) using evaporation of W atoms in a low pressure H$_2$S atmosphere.  It is characterized by means of scanning tunneling microscopy, low-energy electron diffraction and core-level spectroscopy. 
Its electronic band structure is determined by angle-resolved photoemission spectroscopy. The valence band maximum at $\bar{K}$ is found to be significantly higher than at $\bar{\Gamma}$. The observed dispersion around $\bar{K}$  is in good agreement with density functional theory calculations for a free-standing monolayer, whereas the bands at $\bar{\Gamma}$ are found to be hybridized with states originating from the Au substrate. Strong spin-orbit coupling leads to a large spin-splitting of the bands in the neighborhood of the $\bar{K}$ points, with a maximum splitting of 419(11)~meV. The valence band dispersion around $\bar{K}$ is found to be highly anisotropic with spin-branch dependent effective hole masses of $0.40(02)m_e$ and $0.57(09)m_e$ for the upper and lower split valence band, respectively. The large size of the spin-splitting and the low effective mass of the valence band maximum make single-layer WS$_2$ a promising alternative to the widely studied MoS$_2$ for applications in electronics, spintronics and valleytronics.   
\end{abstract}



\maketitle

\section{Introduction}

The discovery of graphene \cite{novoselovelectric2004,Zhang:2005ab,Novoselov:2005aa} established the possibility to obtain stable two-dimensional solids, and it was soon realized that layered materials other than graphite can be used as bulk parents for novel two-dimensional materials \cite{novoselovtwo-dimensional2005}. As in graphite, the weak van der Waals interactions between adjacent atomic layers in alternative parent materials, such as transition metal dichalogenides (TMDCs), permit mechanical exfoliation of a single-layer (SL) with electronic properties that differ in subtle but important ways from those of the bulk material. One example is the indirect to direct band gap transition observed in case of MoS$_{2}$ \cite{makatomically2010,Splendiani:2010aa} and other TMDCs \cite{zhangdirect2013}.
 
SL TMDCs consist of a layer of metal sandwiched between two layers of chalcogens. These materials exhibit a variety of electronic properties, ranging from insulating to metallic \cite{chhowallathe2013}. Probably the most studied SL TMDCs are molybdenum dichalcogenides, which are attractive materials for electronic applications such as transistors \cite{radisavljevicsingle-layer2011,krasnozhonmos22014}, diodes \cite{Lin:2015}, photoemitting devices \cite{sundaramelectroluminescence2013}, solar cells \cite{bernardiextraordinary2013} and memristors \cite{sangwangate-tunable2015}. Furthermore, the unique spin texture of both conduction and valence band makes SL TMDCs well-suited for studying quantum degrees of freedom such as spin or valley pseudospin or their interactions \cite{shanoptical2015,wuelectrical2013,xuspin2014}. In the case of the tungsten dichalcogenides, much stronger spin-orbit coupling is expected than in the case of the Mo-based analogues, and the properties just enumerated  should thus be more stable at room temperature \cite{zhugiant2011, Yeh:2015, Latzke:2015}. Naively, a strong spin-orbit splitting of the bands can be expected to result in an increased band curvature near the top of the valence band and hence to a reduced effective mass; and indeed, WS$_2$ is predicted to be the best material among all of the TMDCs for a transistor channel, due to its low effective hole mass \cite{ovchinnikovelectrical2014,liuperformance2011, kormanyos2015}.

Many proposed approaches to characterizing the electronic properties of SL TMDCs, as well as many potential applications, require large area and high quality samples. In the present work, we introduce a method for the epitaxial growth of WS$_2$ on Au(111). We study the growth and structure by scanning tunneling microscopy (STM), low energy electron diffraction (LEED) and core level spectroscopy. The high quality of the SL WS$_2$ obtained in this procedure permits an investigation of the electronic structure by angle-resolved photoemission spectroscopy (ARPES). The results confirm the expected strong splitting of the SL WS$_2$ valence band at $\bar{K}$, indications of which had already been seen in early experiments on single layers \cite{Klein:2001}. The splitting is found to be almost three times larger than the spin-splitting determined for SL MoS$_2$ \cite{miwaelectronic2015} with a strong warping of the constant energy contours around the $\bar{K}$ point. The expected lower effective mass near the valence band maximum is also confirmed.

\section{Methods}

Growth and measurements were performed at the SGM3 endstation at  the ASTRID2 synchrotron radiation facility \cite{hoffmannan2004}. Prior to growth, the Au(111) single crystal substrate was thoroughly cleaned by repeated cycles of Ne ion sputtering (E=0.75~keV) and annealing (600$^{\circ}$C) in ultra high vacuum (UHV) until the regular herringbone reconstruction was observed by STM. 
 In the first step of the growth procedure, tungsten was evaporated onto the clean surface using a commercial e-beam evaporator charged with a 99.95\% purity W rod. During the evaporation, the sample was exposed to 99.6\% purity H$_2$S using a homemade nozzle situated $\approx 1$~cm from the sample. This allowed for a local pressure at the sample face that was higher than the background pressure in the chamber (which was maintained at $\approx 10^{-6}$~mbar). In the second step, the sample was annealed to 650$^{\circ}$C for 30 minutes, maintaining the H${_2}$S atmosphere. This procedure is similar to that used in Ref. \onlinecite{fuchtbauermorphology2013} to obtain WS$_2$ nano clusters. In order to obtain a large coverage of WS$_2$, however, it is necessary to repeat the above sequence. The number of growth cycles determines the coverage of the SL WS$_2$ film. A last anneal at 750$^{\circ}$C under UHV conditions was used to desorb any residual contamination from the growth process. The sample quality was examined by STM, LEED and core level spectroscopy using a photon energy of 140~eV. The reported core level binding energies have been calibrated using the bulk component binding energy of the clean Au(111) substrate \cite{Heimann:1981aa}.  The characterization of the film was performed at room temperature.  
 
ARPES measurements were carried out at a temperature of $\approx 110$~K, with energy and angular resolutions better than 20~meV and 0.2$^{\circ}$, respectively. Measurements were taken with photon energies in the range of 14-80~eV. The WS$_2$-related bands did not show a $k_{z}$ dispersion in this energy range, confirming the two-dimensional character of the system (photon energy scans not presented here). 

\section{Results and Discussion}

\begin{figure}
	\includegraphics[width=8.2cm]{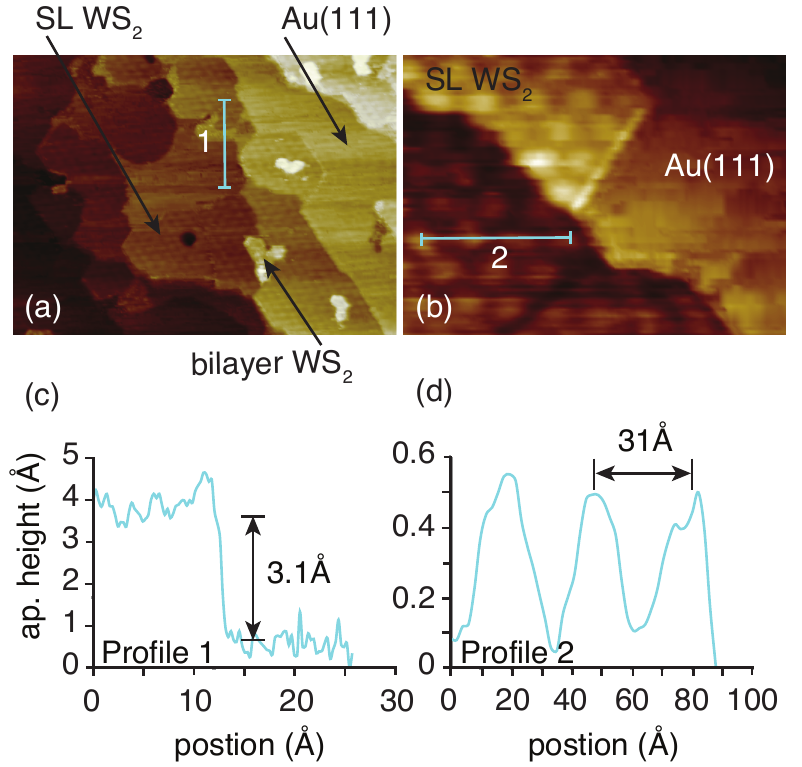}%
	\caption{(color online) (a) Large-area STM image, showing regions of bare Au(111), SL WS${_2}$ (coverage of $\approx$0.7~ML) and bilayer WS${_2}$. (b) Small-area  image emphasizing the moiré structure which has a measured   periodicity of $\approx 31$~\AA. (c) Apparent height profile along line 1 in panel (a), showing an approximate height of $\approx 3.1$~\AA~ for  SL WS${_2}$. (d) Line profile acquired along line 2 in panel (b) showing the  moiré periodicity.    Scanning parameters : (0.06~V, 0.2~nA) and (0.62~V, 0.28~nA), in (a) and (b) respectively.  The STM images were analyzed using the free WSxM software \cite{Horcas:2007}.}
	\label{fig:fig1}
	
\end{figure}

Representative STM images of SL WS$_2$ on Au(111) are shown in Fig.~\ref{fig:fig1}. The small lattice mismatch between the substrate ($a_{Au(111)}=2.88$~\AA) and SL WS$_2$ ($a_{WS_2}=3.15$~\AA) \cite{fuchtbauermorphology2013} leads to a modulation of the local density of states, resulting in the observed moiré pattern (Fig.~\ref{fig:fig1}(b)). The apparent height of the WS$_2$ ($\approx 3.1$~\AA) and the moiré superstructure periodicity ($\approx 31$ ~\AA) are determined from the respective line profiles in Fig.~\ref{fig:fig1}(c) and (d), and are consistent with the results for WS${_2}$ and MoS${_2}$ nano-islands  \cite{fuchtbauermorphology2013,sorensenstructure2014}. The SL WS$_2$ coverage of the sample shown in Fig. \ref{fig:fig1} is approximately 0.7 monolayers (ML). At this coverage, some bilayer regions begin to be observed.  This amount of bilayer coverage is not large enough to be detectable in ARPES, where the formation of a bonding / anti-bonding splitting of the topmost valence band leads to a distinct difference between the electronic structure of bilayer and SL TMDCs  \cite{cheiwchanchamnangijquasiparticle2012} (see further discussions of this below).
 
\begin{figure}
	\includegraphics[width=8.2cm]{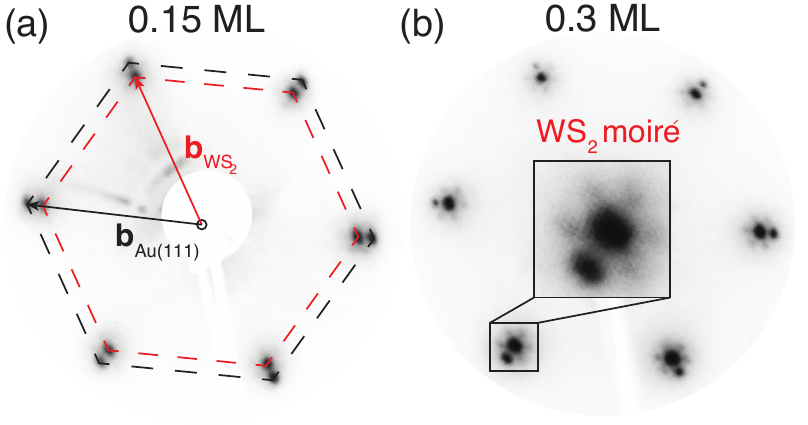}%
	\caption{(color online) LEED images  at kinetic energy of 118~eV for samples of different coverages. (a) For a low coverage sample ($\approx 0.15$~ML)  two concentric hexagonal patterns are observed originating from the substrate and SL WS${_2}$. The reciprocal unit cell vectors are indicated with arrows. (b) In case of samples with higher coverage ($\approx 0.3$~ML) a moiré pattern is observed due to sufficiently large  island size. The inset presents a magnified moiré pattern observed around one of the main spots.}
		\label{fig:fig2}
	\end{figure}

The formation of the moiré pattern can be followed by LEED measurements taken at lower coverages as shown in Fig.~\ref{fig:fig2}. For a coverage of $\approx 0.15$~ML, two concentric hexagonal patterns dominate (Fig.~\ref{fig:fig2}(a)). Based on length of the unit cell vectors, the outer and inner hexagon can be assigned to Au(111) and SL WS$_2$, respectively. The measured reciprocal unit cell vector ratio $b_{Au(111)}$/$b_{WS_{2}}$=1.1 is in excellent agreement with the crystal unit cell vector ratio $a_{WS_{2}}$/$a_{Au(111)}$=1.1. In the case of the higher coverage ($\approx 0.3$~ML) shown in Fig.~\ref{fig:fig2}(b), the moiré pattern is clearly observed in addition to the spots caused by the separate reciprocal lattice vectors of Au(111) and WS$_2$. The presence of the moiré pattern is consistent with the STM results reported in Fig.~\ref{fig:fig1}(b). Its absence for the lowest coverage can be explained by the fact that the growth starts with the formation of small islands \cite{fuchtbauermorphology2013}, which are initially too small to establish a moiré pattern. Similar LEED patterns have also been observed for graphene grown on transition metals \cite{NDiaye:2008aa}.

\begin{figure}
	\includegraphics[width=6.2cm]{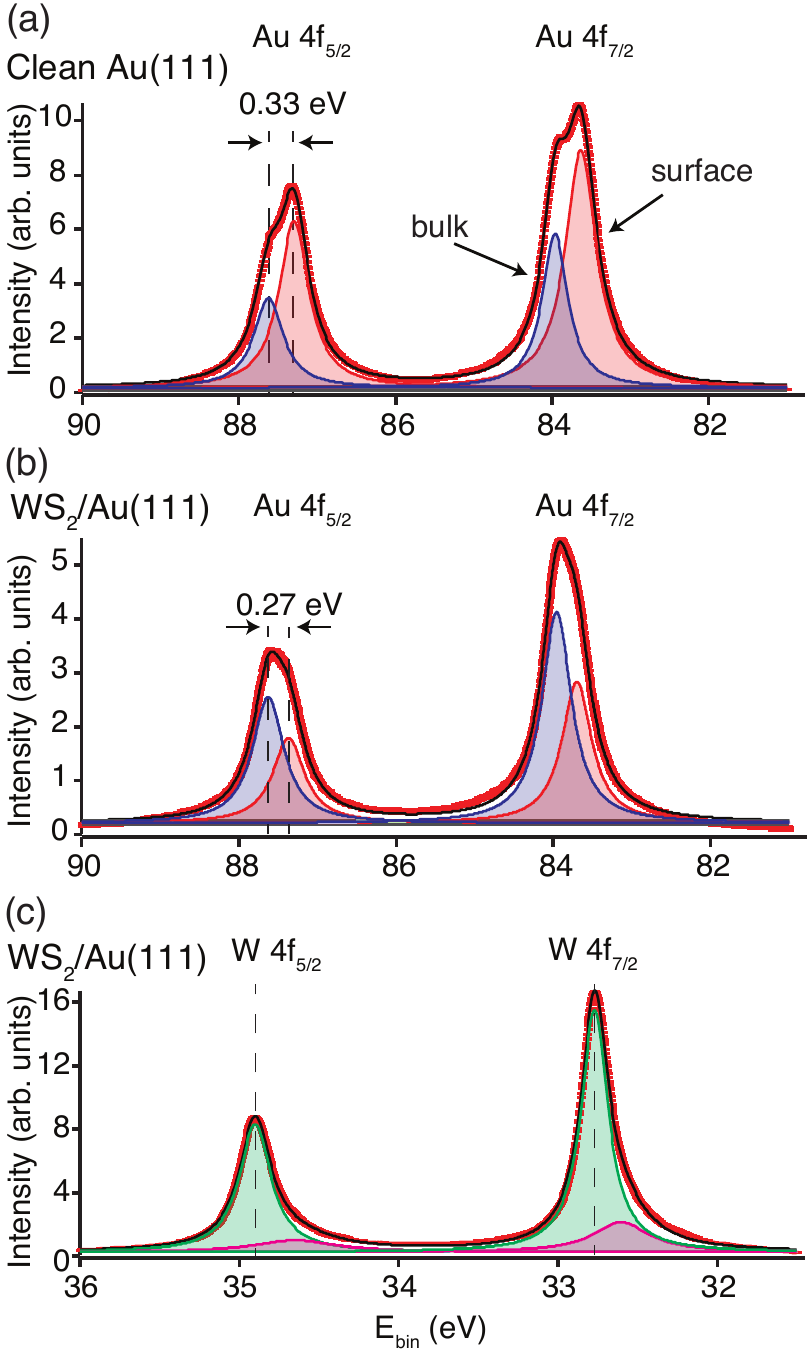}%
	\caption{(color online) (a-b) Au$_{4f}$ core levels for clean Au(111) and $\approx$0.7\,ML of SL WS$_2$/Au(111), respectively. Blue and red peaks correspond to bulk and surface components, respectively. (c) W$_{4f}$ core levels from SL WS${_2}$. Small peaks observed at lower binding energy (in magenta) indicate the presence of W atoms with lower oxidation states (WS$_{2-x}$).}
	\label{fig:fig3}
\end{figure}

Fig. \ref{fig:fig3} shows angle-integrated core level spectra taken in order to follow the local chemical changes during the growth of  SL WS${_2}$/Au(111). All spectra were analyzed by the subtraction of a Tougaard-type background \cite{tougaardquantitative1988} and subsequently fitted by Lorentzian profiles~\cite{Heimann:1981aa}. Before the growth, the Au~4f core levels of the clean Au(111) crystal were measured as a reference (Fig.~\ref{fig:fig3}(a)). A fit reveals a bulk component and a surface-shifted component at lower binding energy, in good agreement with the literature\cite{Heimann:1981aa}.   The growth of approximately 0.7\,ML of SL WS$_2$ was found to affect the Au~4f core levels as shown in Fig. \ref{fig:fig3}(b).  The surface component intensity is largely reduced and shifted towards the bulk component by 58(32)~meV with respect to clean Au(111). It is difficult to analyze this spectral change in detail, as there are several factors contributing to it. The first is the actual suppression or shift of the surface core level component due to the interaction with the sulphur. The second is the remaining presence of clean Au(111) patches that should give rise to a much weaker but largely un-shifted surface component.

The W~4f core level spectrum consists of a spin-orbit split doublet with each peak showing an intense high binding energy component and a weak low energy component.  The observed binding energies for the main components are E$_{5/2}=34.85(02)$~eV  and E$_{7/2}=32.72(01)$~eV. The spin-orbit splitting of the states is thus 2.13(02)~eV. These values are consistent with measurements on other WS$_2$ systems \cite{shpakxps2010,fuchtbauermorphology2013, Cattelan:2015}. However, there is a clear difference between the spectra here and those previously reported for smaller nano-scale WS$_2$ islands on Au(111) \cite{fuchtbauermorphology2013}, for which two almost equally strong components were found in each spin-orbit split peak of the W 4f core level. Those findings were interpreted in terms of core level components from the edges as well as from the basal plane of the WS$_2$ nano clusters, consistent with findings for MoS$_2$ clusters \cite{Bruix:0aa}. This interpretation is supported by our result of only one main component, as the number of edge atoms with lower coordination is small for the high coverage realized here. The weak low binding energy component observed in Fig.~\ref{fig:fig3}(c) might stem from edge atoms, or from the presence of W atoms in lower oxidation states, probably due to not fully sulfidized regions WS$_{2-x}$ \cite{shpakxps2010,fuchtbauermorphology2013}. The presence of metallic W(0) or W(+II) is not observed. 

\begin{figure*}
		\includegraphics[width=1\textwidth]{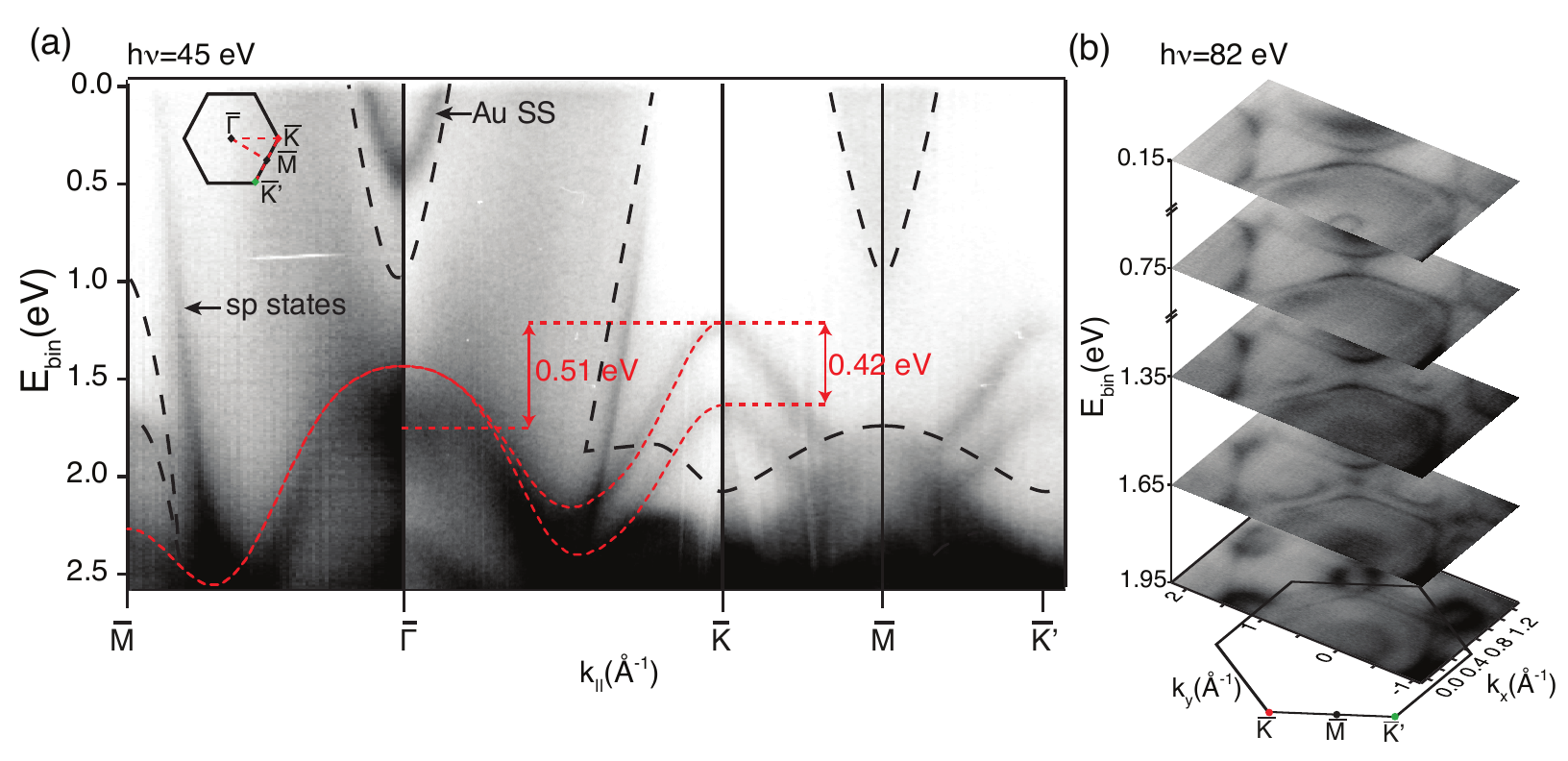}
		\caption{(color online) (a) ARPES spectra of SL WS${_2}$ on Au(111) along high-symmetry directions shown in the inset. The well-defined contribution from the Au surface state (SS) and the $sp$ bands are indicated in the image. The projected bulk band gaps of  Au(111)  \cite{Takeuchi:1991} are outlined by dashed black lines. The energy differences between the valence band local extrema at the $\bar{\Gamma}$ and $\bar{K}$ (0.51~eV) points and spin-splitting of the bands at $\bar{K}$ (0.42~eV) are marked. The theoretical dispersion for free-standing SL (red dashed line) is superimposed on the data (after Ref. \onlinecite{zhugiant2011}). (b) Photoemission intensity along constant energy contours throughout the surface Brillouin zone at different binding energies.
			\label{fig:fig4}}
\end{figure*}

Fig.~\ref{fig:fig4}(a) shows the electronic structure of  SL WS$_2$/Au(111) along different high symmetry directions of the Brillouin zone as observed by ARPES. Corresponding constant energy contours are presented in Fig.~\ref{fig:fig4}(b). Sharp features attributed to Au(111), such as the surface state \cite{lashellspin1996}, $sp$ bulk bands and projected bulk gap edges \cite{Takeuchi:1991}, are marked in Fig.~\ref{fig:fig4}(a).  

The uppermost valence band of SL WS$_2$ is discernible for binding energies larger than 1.2~eV, especially between $\bar{\Gamma}$ and $\bar{K}$ and towards $\bar{M}$. The band is significantly sharper within the projected bulk band gaps of Au around $\bar{K}$ (outlined by dashed lines) than near $\bar{\Gamma}$ and $\bar{M}$. Near $\bar{M}$ the WS$_2$ states strongly hybridize with the gold continuum and cannot be discerned. The maximum of the valence band is found to be situated at $\bar{K}$ ($\approx 0.51$~eV higher than at $\bar{\Gamma}$), consistent with the expected direct band gap at $\bar{K}$. The observation of a single valence band at  $\bar{\Gamma}$ with a higher binding energy than at $\bar{K}$ also rules out a significant contribution from bilayer WS$_2$: in bilayer TMDC systems, the valence band near $\bar{\Gamma}$ shows a bonding / anti-bonding splitting that would be observable by ARPES as two bands, in contrast to what is seen here \cite{zhangdirect2013}. 

The overall observed dispersion of the band structure is in good agreement with calculations for a free-standing SL WS$_2$  (red dashed lines)\cite{zhugiant2011}, notably around $\bar{K}$. Divergence from these calculations in terms of a shift towards higher binding energy is seen near $\bar{\Gamma}$, similar to what has previously been observed for SL MoS$_2$/Au(111), where it was attributed to an interaction with the substrate \cite{miwaelectronic2015}. The bands around $\bar{\Gamma}$ are a mixture of out-of-plane W d$_{z^2}$ and S p$_{z}$ orbitals, which can participate in bond formations with the Au and as a result change the band dispersion \cite{zhugiant2011}. In contrast to this, the valence bands at the $\bar{K}$ points are derived mainly from in-plane W d$_{xy}$ and d$_{x^2-y^2}$ orbitals. These are not only less likely to be affected by the interaction with the substrate, but the states nearby are also situated in a projected bulk band gap. Note that this situation is distinctly different from the case of graphene where the bands near $\bar{K}$  are formed from out-of plane $\pi$ orbitals. This difference in orbital character near the $\bar{K}$ point also explains the absence of replicas induced by the moiré superstructure for SL WS$_2$/Au(111), something that is observed in ARPES from epitaxial graphene systems \cite{pletikosicdirac2009,kraljgraphene2011}. 

Perhaps the most interesting feature of the SL WS$_{2}$ valence band structure is the large spin-splitting at the $\bar{K}$ points. This lifting of the spin degeneracy is a characteristic feature of the SL and due to the removal of the inversion symmetry present in the 2H bulk material and in the bilayer \cite{cheiwchanchamnangijquasiparticle2012,rileydirect2014,suzukivalley-dependent2014}. However, even in the SL material, the splitting is constrained by time-reversal symmetry. This forbids a splitting at $\bar{\Gamma}$ and the additional combination with translational symmetry also forbids a splitting at $\bar{M}$. In a simple picture, the size of the splitting at $\bar{K}$ strongly depends on the atomic spin-orbit splitting and it is thus expected to be significantly larger for W than for Mo.  

\begin{figure}
	\includegraphics[width=8.2cm]{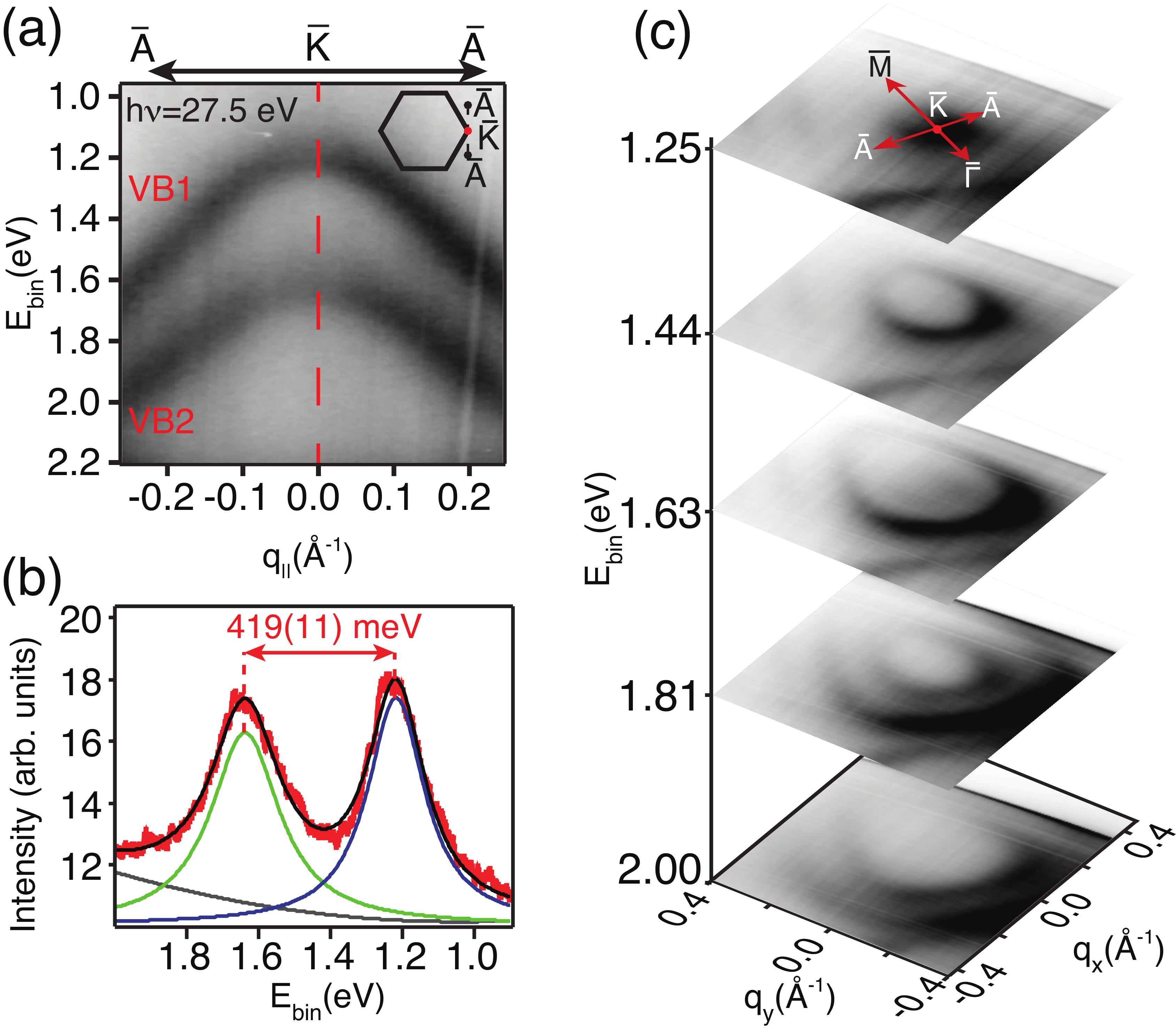}
	\caption{(color online) (a)ARPES spectrum of SL WS${_2}$ on Au(111) along the $\bar{A}$-$\bar{K}$-$\bar{A}$ direction, as indicated in the inset.(b) Energy distribution curve through the $\bar{K}$ point (red dashed line in (a)) fitted with two Lorentzian profiles and quadratic background. (c) Constant energy contours close to the $\bar{K}$ point. The $q_x$ and $q_y$ coordinates have $\bar{K}$ as origin.}
	\label{fig:fig5}
\end{figure}

The spin-splitting near $\bar{K}$ is investigated in more detail using the data in Fig.~\ref{fig:fig5}, which shows a spin-splitting of the valence band extrema of $\Delta E_{VB}$=419(11)~meV (Fig.~\ref{fig:fig5}(b)). This does indeed greatly exceed the value observed for SL MoS$_{2}$/Au(111) ($\Delta E_{VB}$=145(4)~meV) \cite{miwaelectronic2015}, and is in good agreement with theoretical predictions \cite{Ramasubramaniam:2012} and with values measured for analogous exfoliated materials \cite{Zhao:2013}.  Among the layered TMDCs, only SL WSe$_{2}$ ($\Delta E_{VB}$=462~meV) and SL WTe$_{2}$ ($\Delta E_{VB}$=480~meV) are expected to exhibit a larger spin-splitting \cite{zhugiant2011,kormanyos2015}, something that was recently confirmed for SL WSe$_{2}$ \cite{lespinorbit2015}. 

A closer look at the contours in the constant energy surfaces (Fig.~\ref{fig:fig5}(c)) reveals an anisotropy of the valence band dispersion around $\bar{K}$, visible as a trigonal warping (TW). This effect is theoretically predicted and in the simplest approximation can be described as a third order correction to the parabolic energy dispersion~\cite{kormanyos2015}. In the  $\mathbf{k}\cdot\mathbf{p}$ formalism, it is caused by the interaction terms between the uppermost valence band with the lower lying valence bands \cite{zhugiant2011,kormanyosmonolayer2013}. TW reflects the underlying three-fold rotational symmetry of the crystal structure. It should be noted that this effect is connected to the general electronic structure rather than to the relativistic  spin-orbit coupling: thus, for example, it is also present for graphite and graphene where spin-orbit coupling is negligible \cite{Shirley:1995aa,mucha-kruczynskicharacterization2008}.

Given the constraint that the spin-splitting has to vanish at $\bar{\Gamma}$ and $\bar{M}$, the size of the splitting at $\bar{K}$ can be expected to indirectly affect the band curvature of the valence band maximum and thus the hole effective mass of the material. This is investigated by determining the effective masses near $\bar{K}$. For consistency with calculations \cite{kormanyos2015} and to avoid the problem of a directional dependence due to the TW, the effective mass has been fitted in a region very close to the $\bar{K}$ point ($\approx 0.07$~\AA$^{-1}$) along the $\bar{\Gamma}$-$\bar{K}$-$\bar{M}$ direction. This procedure leads to hole effective masses for VB1 and VB2 (defined in Fig.~\ref{fig:fig5}(a)) as $m_{VB1}=0.40(02)m_e$ and $m_{VB2}=0.57(09)m_e$. The same fitting procedure applied for SL MoS$_2$/Au(111)  \cite{miwaelectronic2015} yields $m_{VB1}=0.55(03)m_e$ and $m_{VB2}=0.67(04)m_e$. The effective mass of the uppermost valence band is therefore indeed reduced, consistent with the naive expectation (even though this model does not attempt to explain why $m_{VB2}$ is also reduced). All these values are in good agreement with calculations \cite{kormanyos2015}.  

Finally, an interesting difference between the data presented here and that reported previously for SL MoS$_2$/Au(111) \cite{miwaelectronic2015} is the very clear presence of the Au(111) surface state near $\bar{\Gamma}$ seen in Fig.~\ref{fig:fig4}. For SL MoS$_2$/Au(111), only a very faint signature of this state was reported. This difference is ascribed to the slightly different preparation procedure. In the present work, the synthesis and all the analysis were performed in a single UHV system without ever exposing the sample to air. In the previous work for SL MoS$_2$/Au(111), synthesis and ARPES were performed in separate UHV systems, and the sample was transferred between them through air. After such a transfer, atomically clean SL MoS$_2$ can be recovered by a brief anneal, and the measured electronic structure of the layer is not affected. This, however, is not necessarily the case for the remaining clean Au(111) terraces, where the surface state might remain quenched by adsorbed contaminants --- at least, this may be the case for the low annealing temperature of 500~K used in Ref. \onlinecite{miwaelectronic2015}.  The observed surface state in Fig.~\ref{fig:fig4} is thus likely to be located on the remaining clean terraces and not under the SL WS$_2$.

\section{Conclusions}

We have introduced a preparation method for high-quality epitaxial SL WS$_2$ on Au(111) and studied the electronic structure of this system. The observed valence band dispersion of SL WS$_2$/Au(111) around $\bar{K}$ and the spin-splitting at $\bar{K}$ of $\Delta E_{VB}$=419(11)~meV are found to be in good agreement with calculations for the free-standing layer \cite{zhugiant2011}. The strong spin-orbit splitting contributes to a lowering of the hole effective mass near the valence band maximum, suggesting that WS$_2$ could be a more suitable material for electronic applications than MoS$_2$.

\begin{acknowledgments}  
We gratefully acknowledge financial support from the VILLUM foundation, the Danish Council for Independent Research, Natural Sciences under the Sapere Aude program (Grant No. DFF-4002-00029), the Lundbeck Foundation, the Danish Strategic Research Council (CAT-C) and Haldor Tops\o e A/S.  
\end{acknowledgments}

%

\end{document}